\documentclass[runningheads]{llncs}
\usepackage{graphicx}
\graphicspath{ {figures/} }

\usepackage{array}
\usepackage{multirow}

\usepackage{arydshln}

\usepackage{mathtools} 
\usepackage{amsmath}

\begin{document}
\title{An Experimental and Comparative Benchmark Study Examining Resource Utilization in Managed Hadoop Context}
\titlerunning{Benchmark: Utilization in Managed Hadoop Context}
\author{Uluer Emre Özdil\inst{1}\orcidID{0000-0002-7715-4652} \and
Serkan Ayvaz\inst{2}\orcidID{0000-0003-2016-4443 } }
\authorrunning{U.E.Özdil and S.Ayvaz}
\institute{Big Data Analytics and Management, Bahcesehir University, Istanbul, Turkey \and
Dept. of Computer Engineering, Yildiz Technical University, Istanbul, Turkey
\email{sayvaz@yildiz.edu.tr}\\
%\url{http://www.springer.com/gp/computer-science/lncs}
}
\maketitle              % typeset the header of the contribution
\begin{abstract}
Transitioning cloud-based Hadoop from IaaS to PaaS, which are commercially conceptualized as pay-as-you-go or pay-per-use, often reduces the associated system costs. However, managed Hadoop systems do present a black-box behavior to the end-users who cannot be clear on the inner performance dynamics, hence, on the benefits of leveraging them. In the study, we aimed to understand managed Hadoop context in terms of resource utilization. We utilized three experimental Hadoop-on-PaaS proposals as they come out-of-the-box and conducted Hadoop-specific workloads of the HiBench Benchmark Suite. During the benchmark executions, we collected system resource utilization data on the worker nodes. The results indicated that the same property specifications among cloud services do not guarantee nearby performance outputs, nor consistent results within themselves. We assume that the managed systems' architectures and pre-configurations play a significant role in the performance.

\keywords{Big Data \and Managed Hadoop \and Hadoop-on-PaaS\and HiBench \and Performance evaluation}
\end{abstract}
\section{Introduction}
Big Data is an indispensable area for both enterprises and academia in the information era. The term refers to datasets associated with overwhelming size, ever-growing speed, and having a variety of data structures. New approaches have emerged in order to ease the maintenance of Big Data and enable valuable insights from it by leveraging complex statistical formulae.

Frameworks for distributed storage and computation owe a great deal to web indexing engines for inheriting motivation and foundational work done previously by the open-source community. This collective efforts yielded what is known as Hadoop \cite{noauthor_apache_nodate} and its ecosystem today. Considering the complexity of dealing with big data, Hadoop represents a modern analytics framework by decreasing management efforts and analytics operations' duration to an acceptable level by employing affordable commodity computers. 

The commercialization of Cloud Computing in the midst 2000s \cite{noauthor_announcing_nodate} delivered utilization of storage and computing resources to end-users by saving them from high investments in hardware technology that is soon going to be obsolete and is expensive to maintain. As the cloud migration is an ongoing process, Hadoop also slipped out from its on-premise residence to the cloud by being implemented on virtual machine instances, which were provided as IaaS platforms by many providers. 

The Cloud Service Providers (CSP) embraced the need of eliminating Hadoop's complex implementation process on multi-node VMs by providing managed Hadoop systems, commercially presented as PaaS, which are pre-installed and pre-configured Hadoop clusters allowing the installation of tens to hundreds of nodes in a matter of minutes by merely determining some settings like hardware specs and node numbers prior the installation. 

%The managed Hadoop system is both a blessing and a curse. By
Leaving the hard implementation work to a service provider that are often not related to the primary objective of the analysis, the end-users save time and efforts. Nevertheless, this entails a trade-off. By definition, managed systems are prepackaged solutions provided in black-box nature. It means that CSPs apply behind-the-scenes tweaks to achieve better performance on selected approaches like memory-intensive or compute-intensive applications. However, the utilization dynamics of a managed system remains uncertain from the end-user's point of view.

Installing and running a Hadoop cluster comprises intricate processes requiring great amount of investment in terms of hardware, highly skilled technical employees, and most importantly, time. Leveraging tools measuring a big data framework’s performance by previously set evaluation criteria may arise from various motivations. Enterprises would firstly get benchmark reports before approving high-cost investments. From the providers' point-of-view, benchmark reports are strong arguments in terms of marketing if they are good enough or if not, they become signposts for improvement. As for the researchers, benchmarks are useful tools to test different hardware or software constellations and verify optimizations that have been done on respective systems for providing more self-assured results.

In order to deliver trustworthy outputs, benchmarks must be representative and generate realistic workloads to be applied in commonly considered use cases. Big Data frameworks such as Hadoop are complex architectures and provide diverse components including SQL-on-Hadoop, Machine Learning, graph engines, etc. The challenge for benchmark frameworks persists in compensating these requirements. There exist organizations since the invention of relational databases, such as TPC \cite{tpc-history} and SPEC \cite{spec}, focusing on providing industry standard benchmarks and updating their efforts towards cloud computing requirements. Benchmarks are work-in-progress, as surveyed comprehensively in the study of Han et al. \cite{han_benchmarking_2018}, and require constant efforts to stay up-to-date in parallel to developments in big data frameworks.

%[REVİZE: "performans karşılaştırması" ifadesi yerine sistem kaynakları karşılaştırması vurgulandı - emre]

%With the goal of providing a fair comparative performance evaluation of pre-configured managed Hadoop services, the study is an attempt to understand inner dynamics of managed services during benchmark execution, 
The study is an attempt to understand system resource utilization behaviors within pre-configured managed Hadoop services during benchmark execution, address shortcomings of systems' structures and discuss potential reasons. Hence, the main contributions of the study are as follows:

%[REVİZE: çalışmanın yenilikçi yönleri - emre]

\begin{itemize}
% 	\item Hadoop on PaaS are thoroughly investigated as they come out-of-the-box by leveraging workloads from HiBench's Hadoop based benchmark categories Micro, SQL, ML, and Websearch.
% 	\item While conducting HiBench workloads we collected utilization data on the worker nodes within the clusters to enable visual comparison.
% 	\item We inspected Hadoop's execution plans which are kept by HiBench for reference, providing useful explanations especially in cases where there were differences in system utilization, hence, respective performances.

    \item Our evaluation is focusing on the resource utilization in selected clusters with their default configurations rather than comparing performances.
    \item Lifting off the restricting competitive side of benchmark comforted us with the provider selection criterion. We were freed to put a new actor into the sink, bringing diversity to the study.
    \item Our study, by conducting the benchmark in experimental environment, contributes the Managed Hadoop Context itself from the customer's / end-user's point-of-view.
\end{itemize}

\section{Architectural Concepts and Background} 	\label{sec:Background}

Since Big Data emerged as a new field of research, new approaches for facilitating distributed storage and computing paradigms have been proposed as an ongoing maturing process to overcome management and running analytics challenges. In this context, Hadoop has been embraced by a wide spectrum of beneficiaries from industry and academia since its first release in 2005.
In its core functionality, Hadoop centres around the following three concepts:

\begin{itemize}
	\item HDFS filesystem; for storing extensive data across a cluster of nodes,
	\item The MapReduce framework; for distributed computation, and 
	\item YARN; for allocating available resources for the requested tasks.
\end{itemize}

\begin{figure}[!ht]
	\includegraphics[width=\textwidth]{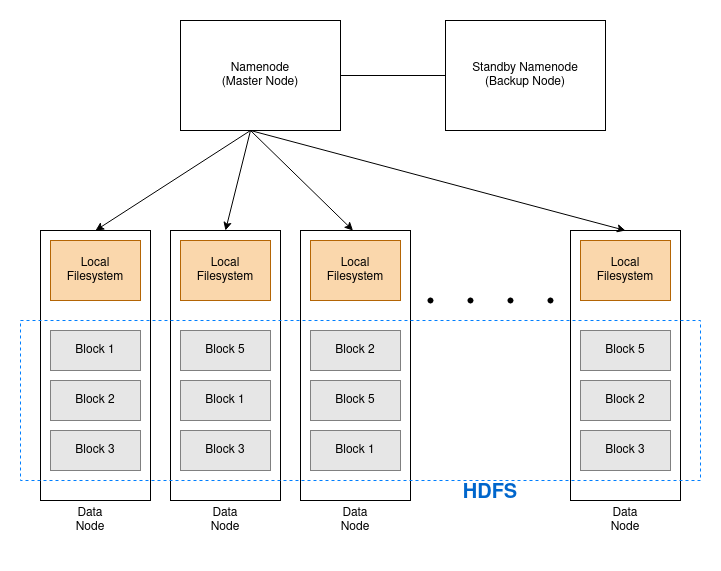}
	\caption{Hadoop Distributed File System}
	\label{fig:HDFSoverview.drawio}
\end{figure}

\subsection{HDFS}\label{HDFS}
Hadoop Distributed File System, which stands for HDFS; is developed with inspiration from the guidelines described in the whitepaper published in 2003 \cite{ghemawat_google_2003} about Google's Filesystem (GFS). GFS is a distributed storage paradigm that can handle petabytes and larger-scale data volumes within a cluster robust to machine failures. 

Significant data volumes are chunked and stored within 128 MB blocks. In HDFS, each block is replicated to different nodes by a factor of 3. These values are the default configurations and can be changed. When the data file is requested, related blocks are constructed from the nodes across the cluster. The redundancy of the blocks guarantees availability. If one or more nodes become out of service, the requested data blocks are gathered from the available redundant copies that are stored on other nodes. 

HDFS is a co-existing file system on the nodes that it is installed. It provides a global distributed view to the files across the cluster, thus listing an HDFS directory is possible from within all nodes. The files on HDFS are listed as they exist on a local filesystem, but the physical parts of the files reside on other physical locations. 

Figure \ref{fig:HDFSoverview.drawio} depicts an overview of HDFS. The architecture of HDFS comprises a Namenode; a dedicated machine to keep track of the files and folders and respective metadata like block locations across the cluster, and many data nodes on which the data blocks are residing \cite{white_hadoop_2015}. Namenode is a single-point-of-failure, meaning if the namenode is down, the whole Hadoop system is down. 

\begin{figure}[!ht]
	\includegraphics[width=0.7\textwidth]{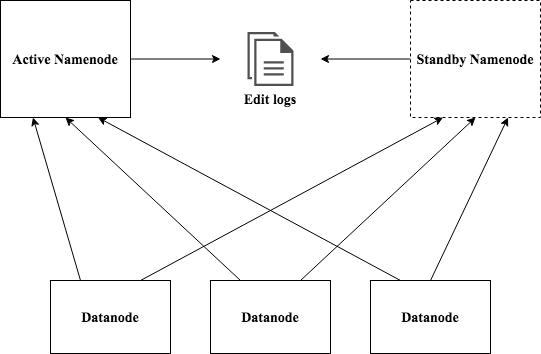}
	\caption{Hadoop High Availability}
	\label{fig:HadoopHA}
\end{figure}

To overcome this issue, starting with version 2, Hadoop matured to a High Availability concept depicted in Figure \ref{fig:HadoopHA} where there exist two namenodes, one active namenode, and one standby namenode communicating with the data nodes and storing edit logs in a shared folder. As the naming refers, the active namenode is in charge, whereas the standby node behaves more like a shadow system. Whenever the active namenode goes down, the standby namenode gets activated so that the services remain available for the end-users.

\subsection{MapReduce}\label{MapReduce}
\begin{figure}[!ht]
	\includegraphics[width=\textwidth]{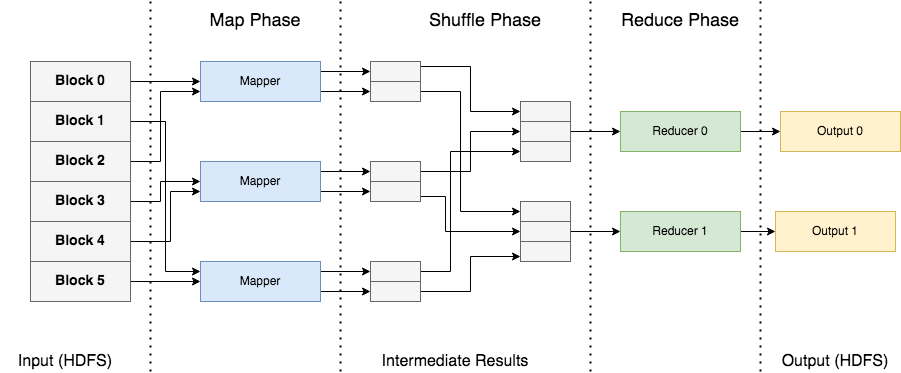}
	\caption{MapReduce execution (Recreated from \cite{schatzle_giant_nodate})}
	\label{fig:MapReduce}
\end{figure}

Similar to HDFS, Hadoop's MapReduce is the open-source implementation of the MapReduce framework described in a Google whitepaper \cite{dean_mapreduce_2004} published in 2004. An MR flow, as depicted in Figure \ref{fig:MapReduce}, starts with assigning blocks from HDFS as input splits to mappers. Computed intermediate results are then shuffled and passed to reducers where outputs are sent back to the client. Nodes which run run mappers and reducers provide parallel processing, and this is scaled by simply adding new commodity computers. 

The computation is made on nodes where related data blocks reside, which explains the term data locality. Executing the computation by leveraging local resources of the node where the respective data block is also stored eliminates the need for moving data across nodes for computation.

\subsection{YARN}\label{YARN}
\begin{figure}[!ht]
	\includegraphics[width=\textwidth]{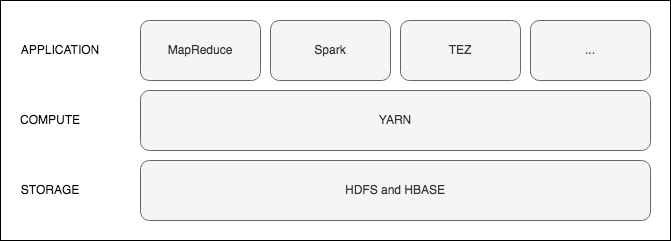}
	\caption{YARN (Recreated from  \cite{white_hadoop_2015}, p. 79)}
	\label{fig:YARNoverview}
\end{figure}

Hadoop version 2.x includes significant architectural improvements in terms of resource management. Version 1.x of Hadoop suffered shortcomings due to an overload of its resource management duties, which was handled within MRv1, where job tracker node and task tracker nodes were running the organizational load of MapReduce executions. YARN standing for "Yet Another Resource Negotiator", emerged as an intermediate layer between HDFS and MapReduce by taking over some of the load that was previously carried out by MRv1, YARN has become a gate to batch and stream operations, interactive queries, and graph processing engines to leverage HDFS file system (Figure \ref{fig:YARNoverview}). 

After YARN, MRv2, which has been stripped from resource managerial duties regarding the previous version, has become more efficient in processing the intended task. YARN innovates Hadoop by bringing the architectural elements such as resource manager, which is one node dedicated to tracking resources across the cluster by availability, and node managers who reside inside each worker node and monitor the containers.

\subsection{Selection of Managed Hadoop Services}\label{selected_csps}Service-level Agreement (SLA) comprising configuration and pricing optimizations are of dynamic nature. This means  improvements on performance or cost may be applied at any time. Thus, favoring some of the services in terms of cost or performance over others would quickly lose ground and become meaningless unless it is not intended by a company for determining which services to choose from.  

%[REVİZE: "performans karşılaştırması" ifadesi yerine sistem kaynakları karşılaştırması vurgulandı - emre]

%The aim of the performance comparison in our study is not to point out the best or worst performing managed proposals across all systems. 
The main focus of the comparison in the study is not based on performances among a variety of managed Hadoop proposals, but the way their worker nodes utilize system resources, which are based on their pre-configured settings. We refrain from commercially favoring any of the providers amongst others, nor do we evaluate their business values. Simply put, the study's objective is to examine the managed Hadoop context in an experimental environment and observe the system utilization behaviors on each cluster in a comparative manner based on the default configurations determined by the provider. We believe this approach makes our findings longer lasting, relevant, and spare from potential impacts of overnight SLA updates. The benchmark results should not be evaluated in the sense of a competition among the service providers since it is not our goal to determine the productivity of the providers over the others.

\begin{figure}[!ht]
	\includegraphics[width=\textwidth]{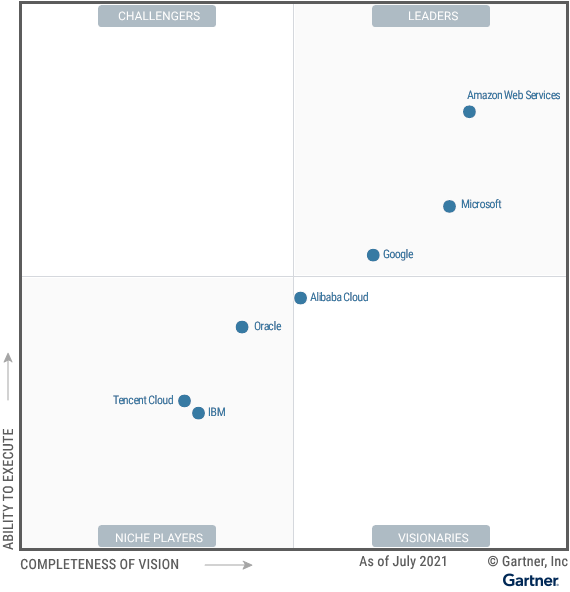}
	\caption{Gartner 2021 - Magic Quadrant for Cloud Infrastructure and Platform Services \textit{Source: \cite{gartner}}}
	\label{fig:gartner2021}
\end{figure}

According to Gartner's 2020 Magic Quadrant for Cloud Infrastructure and Platform Services \cite{gartner}, Amazon Web Services, as the market dominator, is followed by Microsoft Azure and Google Cloud Platform in the leaders section. Alibaba Cloud, Oracle, IBM, and Tencent Cloud reserved their places in the Niche section. Recently, Gartner's 2021 Magic Quadrant for Cloud Infrastructure and Platform Services in Figure \ref{fig:gartner2021}, revised Alibaba Cloud's rank from Niche to the Visionaries section. As for the experimental environment, we focused on the mid-level range in the study. This type of environment appears to be targeted by Microsoft Azure, Google Cloud Platform, and Alibaba Cloud.  

The current market position of the respective benchmark subjects did not play a dominant role in the selection process as our objective was evaluate the resource utilization of managed Hadoop systems rather than their market positioning strategies. Due to the resource limitations in the study, we were required to consider three CSPs only for the evaluations and choose between the options of AWS-EMR and Alibaba Cloud e-MapReduce. Cloud Computing is a fast growing and disruptive technology; we think new players should be taken into account. As there already exist studies evaluating GCP Dataproc, Azure HDInsight and AWS-EMR in a sink as presented in the subsequent Related Work section, and considering its transition potential to a higher level, we decided to include Alibaba Cloud in the evaluations. 

\subsubsection{Microsoft Azure (HDInsight)}\label{Azure_HDInsight}
HDInsight \cite{noauthor_azure_nodate}, Azure's solution to manage Hadoop and Spark platform; is a product of the collaboration with Hortonworks bringing Hortonworks Data Platform (HDP) to a cloud platform \cite{noauthor_azuravail_nodate}. HDInsight differs in architecture to its conjugates by not being a Hadoop cluster installed on the cloud VM service layer, namely Azure Virtual Machines; instead, it is an HDP platform optimized for the cloud. 

Another difference is that HDInsight obligates WASB, the Azure blob system, as the storage system, including an option to leverage Data Lake Storage, excluding Hadoop's native file system HDFS from the options. At the time of the study, Azure was obligating Hadoop to use High Availability mode with two master nodes but leaving hardware selection to the end-user.

\subsubsection{Google Cloud Platform (Dataproc)}\label{GCP_Dataproc}
Google's managed Spark and Hadoop solution, namely Dataproc \cite{noauthor_dataproc_nodate}; is a pre-configured Hadoop on PaaS service built upon pre-installed VM instances on Compute Engine \cite{noauthor_compute_nodate}, which is another service of GCP. An OS spectrum comprising Debian and\textbackslash{or} Ubuntu, which are proposed as pre-installed images during the Dataproc installation process, was offered at the study's date. 

While Hadoop core elements HDFS, YARN, and MapReduce reside as default frameworks, a variety of components related to the Hadoop ecosystem are provided. The end-user is also offered to leverage Cloud Storage \cite{noauthor_cloud_nodate}, Google's proprietary cloud storage service, to store data in the long term as the Dataproc cluster is meant to be terminated after usage. By utilizing Web UI or local CLI via API, the end-users can access the Dataproc cluster and single VM instances within Compute Engine. Google offers a large number of data centers around the globe.

\subsubsection{Alibaba Cloud (e-MapReduce)}\label{Alibaba_eMapReduce}
Alibaba Cloud's managed Hadoop service e-MapReduce \cite{noauthor_what_nodate} leveraging Apache Hadoop and Apache Spark is placed as a service layer on its Elastic Compute Service (ECS) \cite{noauthor_alielastic_nodate}, a similar approach to the one of GCP's. Although Alibaba's number of provided data center locations beyond Mainland China are not as numerous as its conjugates, competitive regions do exist in the United States, Europe, Middle East, and the Asia Pacific. Alibaba Cloud's managed service differs in its pre-installed Operating System. Aliyun Linux 2, which is a Linux distribution based on CentOS, is the open-source version of RedHat Linux developed by Alibaba Cloud. 

AliyunOS proclaims to provide a stable and reliable environment optimized for the Alibaba Cloud infrastructure and be an open source at its GitHub repository \cite{noauthor_alibaba_nodate}. e-MapReduce offers a wide scale of machine types with different specifications for specific purposes like CPU intensive or memory-intensive tasks. As with GCP, Alibaba Cloud e-MapReduce allows the end-users for specifying the type of storage amongst the options of HDFS and Object Storage Service (OSS) \cite{noauthor_oss_nodate}, Alibaba's cloud storage.

\subsection{HiBench Benchmark Suite}\label{benchmarks_hibench}
HiBench is a comprehensive and representative benchmark for Hadoop \cite{noauthor_intel-bigdatahibench_2021}. It was introduced to the open-source community in 2010 \cite{huang_hibench_2010}. It helps evaluating different big data frameworks in terms of speed, throughput, and system utilization. At the date of the study, the current version of HiBench was 7.1 comprising 29 benchmarks which are categorized in 6 sections (Micro, Machine Learning, SQL, Websearch, Graph, and Streaming). Table \ref{tab:hibench-wrkl} lists the Hadoop related workloads that we leveraged in HiBench 7.1 \cite{noauthor_release_nodate}. 

The benchmarks in Micro category were taken from Hadoop's native benchmark tools, where Dfsioe was the enhanced version of its originator Dfsio. It calculates the average I/O rate, averaging throughput of each map task, and aggregating throughput of the cluster. The column "Engine" points out HiBench's dependencies, which are downloaded when Apache Maven compiles the benchmark suite, and leveraged during benchmark executions; Hive engine enables running HiveQL queries for Scan, Join, and Aggregation benchmarks. The Hadoop based ML benchmarks are driven by Mahout. Nutch Engine is utilized for Nutchindexing benchmark. Pegasus, a peta-scale graph mining system, computes PageRank algorithm. The dependencies execute benchmark tasks by translating respective jobs into MapReduce. HiBench's predefined default data scales namely, Tiny, Small, Large, Huge, Gigantic, and Bigdata, provide a scale varying from 32 KB up to 300 GB for Sort, 6 GB for Terasort, and 1.6 TB for Wordcount. User-defined data scales are also applicable within HiBench configurations.

\begin{table}
	\centering
	\small
	\caption{HiBench 7.1 - Hadoop-related Workloads}
	\label{tab:hibench-wrkl}
	\begin{tabular}[b]{ p{3.0cm} p{3.0cm} p{3.0cm} }
		\hline
		\textbf{Category} & \textbf{Workload} & \textbf{Engine} \\
		\hline
		\multirow{4} {3cm} {Micro} & Sort & \multirow{4} {3cm} {M/R} \\
		 & Terasort &  \\
		 & Wordcount & \\
		 & Dfsioe & \\
		\hline
		\multirow{3}{3cm}{SQL} & Scan & \multirow{3} {3cm} {Hive} \\
		 & Join & \\
		 & Aggregation & \\
		\hline
		\multirow{2}{3cm}{ML} & Bayes & \multirow{2} {3cm} {Mahout} \\
		 & Kmeans &  \\
		\hline
		{Websearch} & Pagerank & Pegasus \\
		\hline
	\end{tabular}
\end{table}

\section{Related Work}

Big data emerged as an active area of research recently and has attracted researchers from the open-source community and commercial enterprises from various perspectives. Due high demand for big data tools and solutions, the number of available Big data technologies has been rapidly increasing \cite{ghazal2013bigbench}. As the scales and requirements of institutions greatly differ, the task of choosing most suitable big data systems has become even more challenging. Big data benchmarking plays a key role in evaluating the performances of big data systems for various use case scenarios at different levels \cite{wang2014bigdatabench}. In the review of related work, we focused on the state-of-art Hadoop benchmarking approaches with emphasis on managed Hadoop systems.

Poggi et al. \cite{poggi_characterizing_2018} conducted a comprehensive study for benchmarking Hadoop PaaS services provided by Amazon (EMR), Google (Dataproc), and Azure (HDInsight). The benchmark tool they used, namely BigBench (TPCx-BB), is an industry-standard benchmarking framework developed by Transaction Processing Performance Council (TPC). TPCx-BB comprises 30 business use cases and differentiates from SQL-only benchmarks by also requiring other frameworks like MapReduce, user code (UDF), Natural Language Processing, and Machine Learning. The researchers' study objective was twofold: Firstly, it was characterizing BigBench queries and out-of-the-box performances of Hive and Spark in the cloud, and secondly, it compared the cloud vendors in terms of reliability within a data scale ranging from 1GB to 10TB. The medium-size clusters that were put under test consist of 16 cores and 60+GB memory for the master node and 16 data nodes with 128 total cores. The executions of TCPx-BB along with the data scales showed that Hive performs in most cases better in up to 1TB, whereas, in a 10TB data scale, Spark hits a significantly better performance, up to twice faster than Hive.

In \cite{poggi_state_2016}, Poggi et al. surveyed entry level Hadoop PaaS offerings, namely, Google Cloud Dataproc, Amazon EMR, Azure HDInsight, and Rackspace Cloud Big Data. A local Hadoop cluster was also benchmarked as a criterion setup for comparing the impact of fine-tuning the software stack. Hive was selected as benchmark subject for SQL-on-Hadoop services of the providers, leveraging TPC-H decision support benchmark, where respective cost\textbackslash{}performance rates and scalability in data and computation was considered as well. The study pointed out that the dynamic upgrade capabilities of provider services delivered even better performance after daily based service updates; the pricing varied even with very similar hardware and software specs due to the pricing policies of respective providers.

Wang et al. \cite{wang_improved_2020} utilized HiBench for assessing the effectiveness of their optimization solution on Hadoop's MapReduce framework. The problem definition was targeting the insufficiency of the current MapReduce framework for efficiently handling intermediate data; the output of maps was physically stored on the disk and also read from there to got passed to the reduce slots. As the number of mappers and reducers increase, expensive disk seeks occur which results in high execution times. The researchers' proposal, APA (Aggregation, Partition, and Allocation), aggregated intermediate data in each rack to one file, and a file host pushes the data to reduce tasks. Benchmarking the experimental setup consisting 50 data nodes across 10 racks and 40 Gbps rack interconnectivity yielded 40\% to 50\% performance improvement. 

Hwang et al. \cite{hwang_cloud_2016} used HiBench and four other benchmarks (YCSB, CloudSuite, BenchClouds, and TPC-W) on Amazon EC2 instances introducing new performance metrics applicable to IaaS, PaaS, SaaS, and hybrid clouds. The cloud scaling strategies of vertical scaling (scale-up), horizontal scaling (scale-out), and mixed approach were discussed and evaluated in the study. The researchers conducted the aforementioned benchmarks on cluster scale strategies. The study has drawn multiple conclusions proposing new performance metrics, differentiating between scaling-up and scaling-out strategies and applicable use cases, the performance of scale-out vs. cost-effectiveness of scale-up approaches, and the close relation between cloud productivity and system elasticity, efficiency, and scalability.

Ahn et al. \cite{ahn_performance_2018} put Spark on YARN's performance on the test with HiBench to handle a deluge of data generated by IoT devices. The experiment was run on a cluster with one master and three worker nodes, each node possessing an Intel® Xeon® processor with 20 cores and 128GB main memory meaning 60 cores and 384GB memory in total. HiBench workloads Micro (comprising Sort, TeraSort, and Wordcount), SQL (comprising Aggregation, Join, and Scan), and Machine Learning (comprising Bayes, Logistic Regression, Gradient Boosting Tree, Random Forest, and Linear Regression) was considered with data scale of 30 GB. Spark occupied memory during the whole job execution, which reduces I/Os' negative impact on processor performance. The conductors modified YARN's minimum memory allocation and Spark executor settings to optimize resource usage so that the Spark executors' overall loads remained below total system memory. Alongside with HiBench's duration and throughput report, CPU / memory utilization and disk throughput were profiled as well. This paper's findings pointed out that Spark guarantees performance when provided with enough memory.

Han et al. \cite{han_impact_2017} studied the impact of memory size on big data processing through Hadoop and Spark performance comparison leveraging HiBench's k-Means workload as the only benchmark. For memory sizes 4, 8, and 12 GB, across a data scale from 1 to 8 GB, k-Means benchmark for Hadoop and Spark was executed. The results demonstrated that Spark's overperforming Hadoop unless the total input data size was smaller than 33.5\% of worker nodes' total memory size. After reaching that ratio, Spark suffered from insufficient memory resources. This led to interoperate with HDFS causing a sharp decrease in performance and brings Hadoop to the front in throughput and duration performance. The authors carried out a second experiment to determine if Spark's performance could be improved by tweaking the allocation settings for storage memory and shuffle memory while remaining within the specified memory limitations of 4, 8, and 12 GB. The authors interpreted the benchmark report as an improvement in Spark's processing time, scaling between 5 to 10 percent up to 15 percent.

Ivanov et al. \cite{ivanov_performance_2015} compared the performances of two enterprise-grade applications, DataStax Enterprise (DSE), a production-level implementation of Apache Cassandra with extended features like in-memory computing and advanced security, to name but two, and Cloudera's Distribution of Hadoop (CDH) comprising core Hadoop elements HDFS and YARN integrated with elements belonging to the Hadoop ecosystem. DSE's HDFS compatible file system(CSF) lets Hadoop applications run without any modification. The authors installed the latest stable releases of both software on equal CPU, memory, and network infrastructure configuration; for both installations, default system parameters have been left with their defaults. HiBench's three chosen workloads (CPU-bound wordcount, I/O-bound dfsioe, and mixed HiveBench) were executed three times. Furthermore, the average values have been taken for representativeness. Their study's conclusions proclaimed linear scaling of both systems by increasing data size, while CDH outperformed DSE in read-intensive workloads, DSE performed better in write-intensive workloads. By utilizing HiBench, the study differed from other studies using the YCSB benchmark suite. HiBench's results confirmed the latter's output as well.

Samadi et al. \cite{samadi_performance_2018} conducted an experimental comparison between Spark and Hadoop installed on virtual machines on Amazon EC2 by leveraging nine among the provided HiBench workloads. Accuracy reasons led the conductors to run the workloads three times, concluding input data scales of 1, 3, and 5 GB, respectively. Based on the outputs comprising duration, throughput, speed up, and CPU/memory consumption, it was concluded that Spark consumed less CPU resources and performed better on all workload results over Hadoop. 

\section{Materials and Methods}
%[REVİZE: "performance" ifadesi yerine "utilization" vurgusu getirildi - emre]
%We conducted the benchmark study in two use cases: Use Case 1 aims to map the overall performance behaviors of the services within HiBench's predefined data scales, namely Huge and Gigantic. 
We conducted the benchmark study in two use cases: Use Case 1 aims to cover the overall resource utilization within HiBench's predefined data scales, namely Huge and Gigantic. Workloads from the Hadoop specific benchmark categories (Micro, SQL, ML, and Websearch) were executed. In Use Case 2, we selected two workloads: I/O intensive Sort and CPU intensive Wordcount. The data were scaled up by using HiBench's predefined data scales: Tiny, Small, Large, Huge, and Gigantic. 

Documentation of the Study is publicly available to interested researchers at the GitHub repository \cite{emrettobenchmark}.

\subsection{Experimental Setup}
This section comprises cluster installations, building and running HiBench on installed clusters, capturing system resource utilization data across worker nodes, and the processes to visualize respective data in the study's plots. Detailed documentation for the experimental environment setup and HiBench execution is provided at the reference URL \cite{emretto_setup}.

\subsubsection{Cluster Installations} \label{cluster_installation}

With the goal of benchmarking managed Hadoop with out-of-the-box configurations, we adhered to three principles, given the availability of the options:

\begin{itemize}
    \item the geographic locations of data centers shall be the same or close,
    \item the clusters' computation power, memory and storage capacity shall be same or as close as possible, and 
    \item the Hadoop versions among the offerings shall be same or as close as possible to each other and remain within version limits supported by HiBench 7.1.
\end{itemize}

Thus, Frankfurt was selected as the location for all providers data center. Within given options, we selected 8 CPUs/64GB RAM for the master node and 4 CPUs/32 GB RAM for each worker node totaling in 12 CPUs and 96 GB cluster compute power. 

%[REVİZE: Farklı Hadoop versiyonlarına açıklama getirildi - emre]

Although the selection process with aligned hardware specs was straightforward, the software selection step contained difficulties in finding very same versions of Hadoop. A large variety in different software constellations was provided by all vendors, however, an exact version alignment also covering minor release levels of Hadoop and related software was not possible. As mentioned in Section \ref{YARN}, Hadoop got a major update with version 2, hence, our solution was to pick Hadoop versions with minor differences. After making sure that the releases 2.7.3 \cite{Hadoop2.7.3}, 2.8.5 \cite{Hadoop2.8.5}, and 2.9 \cite{Hadoop2.9.0} are in compliance, these were selected. 

Specified cluster installation options in detail are given in Table \ref{tab:csp-configs}.

\subsubsection{Building and Running HiBench} \label{hibench_installation}

Without applying any performance tweaks to the respective configurations, we immediately executed the benchmarks. We only modified the default configuration in one case, where benchmark was detained from running, reported in Section \ref{discuss}.

HiBench's Hadoop related benchmarks in the categories Micro (Sort, Terasort, Dfsioe, and Wordcount), SQL (Scan, Join, and Aggregation), ML (Bayes and Kmeans), and Websearch (Pagerank) were executed. 

The build and run process in HiBench was straightforward: Downloading the source files, building HiBench with Apache Maven, setting HiBench config files, and running selected benchmarks. The benchmark results comprising data like Duration, Throughput, and respective MapReduce execution plans were stored within the folder "Reports". HiBench was implemented only on the cluster's master node, meaning no additional work was required on the worker nodes.

\begin{table}
	\centering
	\small
	\caption{Selected configurations on CSPs' managed Hadoop services}
	\label{tab:csp-configs}
	\begin{tabular}[!htb]{ p{2.0cm} p{2.3cm} p{3.3cm} p{2.6cm}  }
		\hline
		{} & \textbf{GCP} & \textbf{Azure} & \textbf{Alibaba Cloud}\\
		\hline
		Service & Dataproc & HDInsight & e-MapReduce \\
		Region & europe-west3-a & Germany West Central & eu-central-1 \\
		Location & Frankfurt & Frankfurt & Frankfurt \\
		Image & 1.4-ubuntu18  & HDI 3.6 & EMR-3.32.0 \\
		OS & ubuntu18.04 & ubuntu 16.04 & Aliyun Linux 2 \\
		Hadoop v. & 2.9 & 2.7.3 & 2.8.5 \\
		Java & 1.8.0\_275 & 1.8.0\_275 & 1.8.0\_252 \\
		\hline
		\multicolumn{4}{ c }{MASTER NODE} \\
		\hline
		Machine Type & e2-highmem-8 & A8m v2 & ecs.se1.2xlarge \\
		Processors & 8 vCPU & 8 cores & 8 vCPU \\
		Memory & 64 GB RAM & 64 GB RAM & 64 GB RAM \\
		Immutable Extras & -- & +1 Master Node for High Availability, +3 nodes for Zookeeper & -- \\
		\hline
		\multicolumn{4}{ c }{WORKER NODES} \\
		\hline
		\# of Nodes & 3 & 3 & 3 \\
		Machine Type & e2-highmem-4 & A4m v2 & ecs.se1.xlarge \\
		Processors & 4 vCPU & 4 cores & 4 vCPU \\	
		Memory & 32 GB RAM & 32 GB RAM & 32 GB RAM \\	
		Storage & HDFS 1000 GB & \multirow{3}{ 8em }{WASB \\
			\textit{Azure blob storage}} & HDFS 1000 GB \\	
		Replication & 2 &  & 2 \\	
		Block size & 128 MB &  & 128 MB \\
		\hline
	\end{tabular}
\end{table}

\subsubsection{Capturing Resource Utilization Data} \label{capture_utilization}

By leveraging a low-footprint bash script based on sar commands, we collected system utilization records addressing CPU, Memory, and I/O utilization on each worker node of the cluster during the benchmark execution. The captured data enabled us to visualize the system utilization on each worker node within the cluster.

Due to the small size of the experimental cluster limited to 3 worker nodes, we excluded the network activity track with the assumption that data locality was either provided, or data blocks were handled within rack locality, which was expected to cause the least impact on the network performance.

To capture utilization data, we modified a sar script based on our needs (the original source code was provided by its developer on GitHub \cite{jota_juliojsbsarviewer_2020}). Data capture script was uploaded to each worker node and started manually before the benchmark execution. Subsequent to the benchmark's completion, running scripts on the worker nodes were terminated immediately (Ctrl+C), which resulted in storing the collected raw data to the disk. 

\subsubsection{Visualization} \label{visualization}

The plots are built on the benchmark results and captured cluster utilization data. In order to observe the overall average utilization and how utilization behaves over execution time, we generated two types of plots:

\textit{Spiderplots - For average system utilization:} Spider (radar) plots are useful tools when displaying multiple variables within the same graphic, its use cases mostly address to comparative reasons. With these plots, we aimed to display the average system utilization bound to the benchmark result. A plot comprises six axes, which represent the normalized average values for processor (CPU), Memory (MEM), IO-read and IO-write (IO-R and IO-W, respectively) utilization on the cluster. Benchmark results are represented as Duration (DUR) and Throughput (TPT). Applying a coloring convention enabled differentiation between managed services' utilization results.

The normalization scale was set between 0 and 1, where higher values indicate the better. For the sake of proper visualization, we displayed the shortest Duration as the highest value.

\textit{Linecharts - For system utilization within execution time:} Displaying average values with spider plot drops out the time dimension. Linecharts visualize CPU, Memory, and IO utilization of each worker node of the cluster over a time scale, and enable easy detection of bottlenecks or failures during execution.

The linechart visualizations are organized as follows: Top-left, top-right, and bottom-left plots represent CPU, memory, and I/O utilization on the cluster's each worker node. The x-axis represents the duration of the workload's execution in start time and end time. The left-hand side y-axis measures CPU and Memory usage in percent; the right-hand side y-axis measures I/O-read and I/O-write transfers per second. CPU utilization lines (*cpu\%) are given in blue tones; memory utilization lines (*mem\%) are given in fuchsia tones, I/O-read and I/O-write transfers (*io-r, *io-w) are represented in orange and green tones, respectively. The prefixes w0*, w1*, and w2* refer to the worker nodes to which the representation belongs. The bottom-right plot displays the benchmark results of all three clusters. Duration (shown as an orange bar), which shows the execution time in seconds is better for lower values. Throughput (shown as a green bar) is the amount of processed data in bytes per second. Thus, the higher value of Throughput indicates better performance.

\section{Results} \label{sec:RESULTS}  

Due to space limitations, a small number of visual results are included in this paper. We highly recommend the reader to obtain the pdf file comprising high resolution plots from the Study's documentation repository at the given reference link \cite{emretto_results}.

%[REVİZE: Tablo ve figür referansları eklendi - emre]

The results are provided in two parts: Section \ref{Subsec:RESULTS_OVERVIEW} presents the overall benchmark results in tabular format (Table \ref{tab:uc1-results} and Table \ref{tab:uc2-results}), and averages as spider plots (Figures \ref{fig:uc1-gigantic-new}, Figure \ref{fig:uc2-srt-new}, and Figure \ref{fig:uc2-wrdcnt-new}). We provided \textit{Slowdown estimate} (SE) values in Table \ref{tab:uc1-comparative-results} and Table \ref{tab:uc2-comparative-results}, which show performance differences among the clusters. The performance of Hadoop relies on the number of allocated mappers and reducers, which are specified by a cluster's Hadoop configuration. Therefore,  the numbers of allocated mapper and reducer slots within the clusters are also provided (Table \ref{tab:uc1-sql-mr-allocs} and Table \ref{tab:uc2-mr-allocs}).

%[REVİZE: bu paragrafta performans zayıflığı ifadesi kaynak kullanımında yetersiz kalma bakış açısıyla ifadelendirildi - emre

%In Section \ref{Subsec:RESULTS_ISSUES_BOTTLENECKS} we inspected two cases (the resource utilization for Terasort and Bayes within Use Case 2 for Gigantic data scale). In these benchmarks, a cluster's performance was addressed as failure or performing poorly. To understand if it is due to an architectural issue or a misconfiguration, we leveraged the generated linecharts and MapReduce execution logs, which were stored by HiBench.

In Section \ref{Subsec:RESULTS_ISSUES_BOTTLENECKS} we inspected two cases (Terasort and Bayes within Use Case 2 for Gigantic data scale). In these benchmarks, a cluster's resource utilization was either terminated due to a failure (Figure \ref{fig:uc1-tera-g-cmidt}), or a prolongation in utilization occured which impacted its duration (Figure \ref{fig:uc1-bayes-g-cmidt}). To understand if it is due to an architectural issue or a misconfiguration, we leveraged the generated linecharts and MapReduce execution logs stored by HiBench.

\subsection{Overview}\label{Subsec:RESULTS_OVERVIEW}

Use Case 1 and Use Case 2 raw benchmark results are displayed in Table \ref{tab:uc1-results} and Table \ref{tab:uc2-results} for reference. 

\begin{figure}[!htb]
	\includegraphics[width=\textwidth]{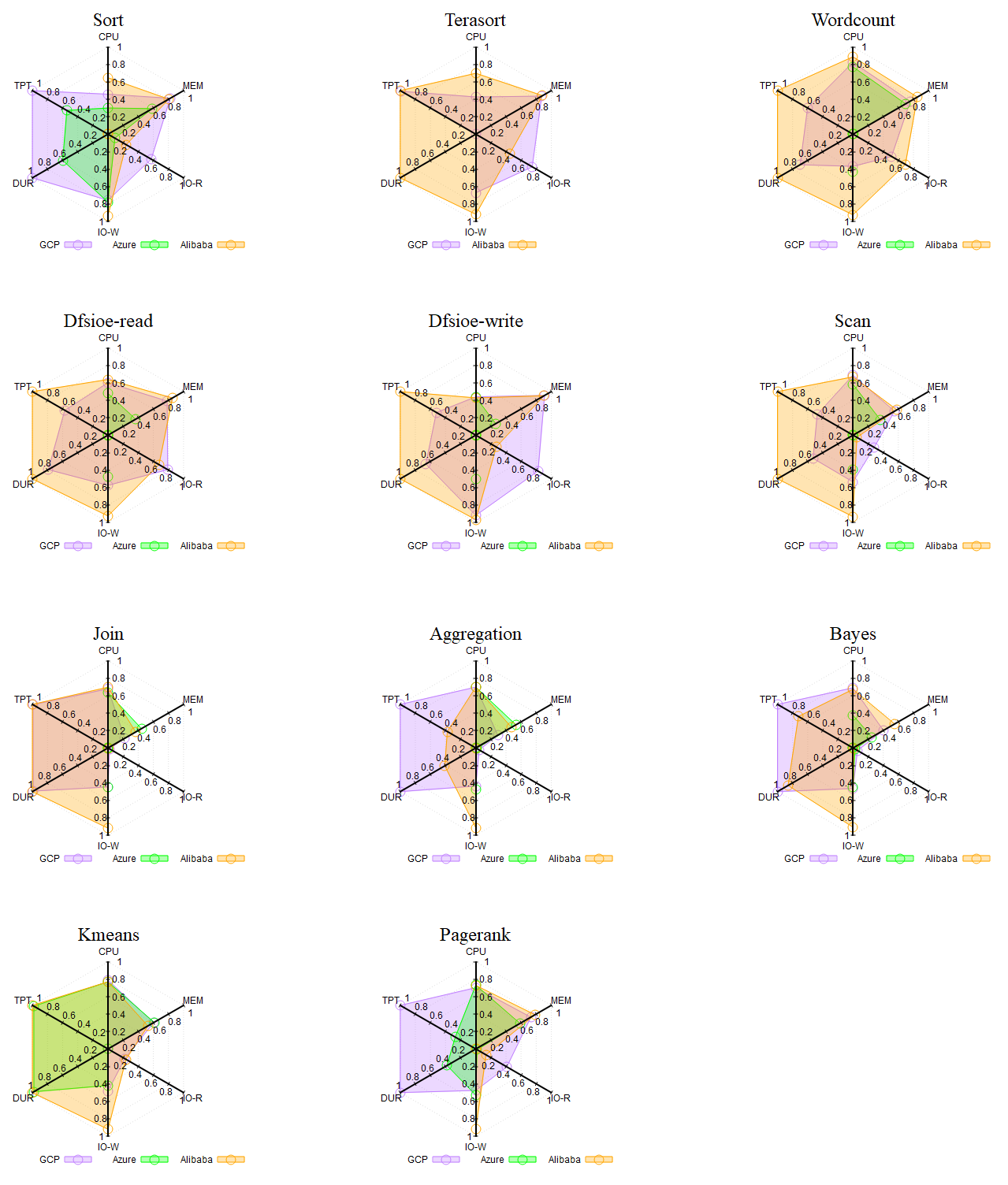}
	\caption{Use Case 1 - Overview to system utilization in data scale Gigantic}\vspace{1em}
	\label{fig:uc1-gigantic-new}
\end{figure}

In Figure \ref{fig:uc1-gigantic-new}, the spider plots depict the main utilization character for Use Case 1 Gigantic data scale, whether it is I/O-bound like in Sort, or CPU-bound as in Wordcount, or both. In most cases, the systems that can utilize more resource displayed the shorter durations. Due to the failed workload execution in Terasort, Azure's Duration and Throughput results could not be provided in this figure.

%[REVİZE: Azure için "performing less than its counterparts" yerine ]

\begin{figure}[!htb]
	\includegraphics[width=\textwidth]{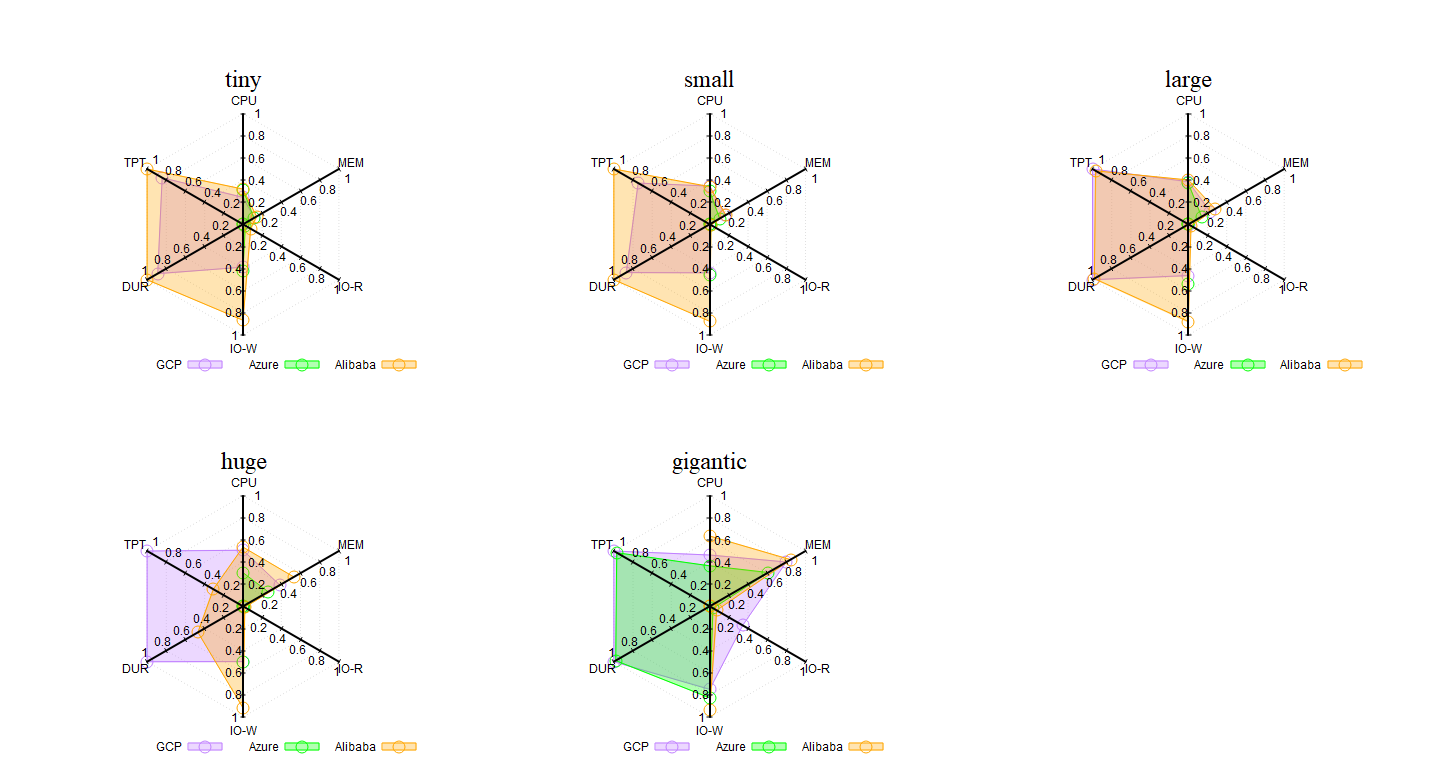}
	\caption{Use Case 2 - Sort performances along with data scales}
	\label{fig:uc2-srt-new}
\end{figure}

\begin{figure}[!htb]
 	\includegraphics[width=\textwidth]{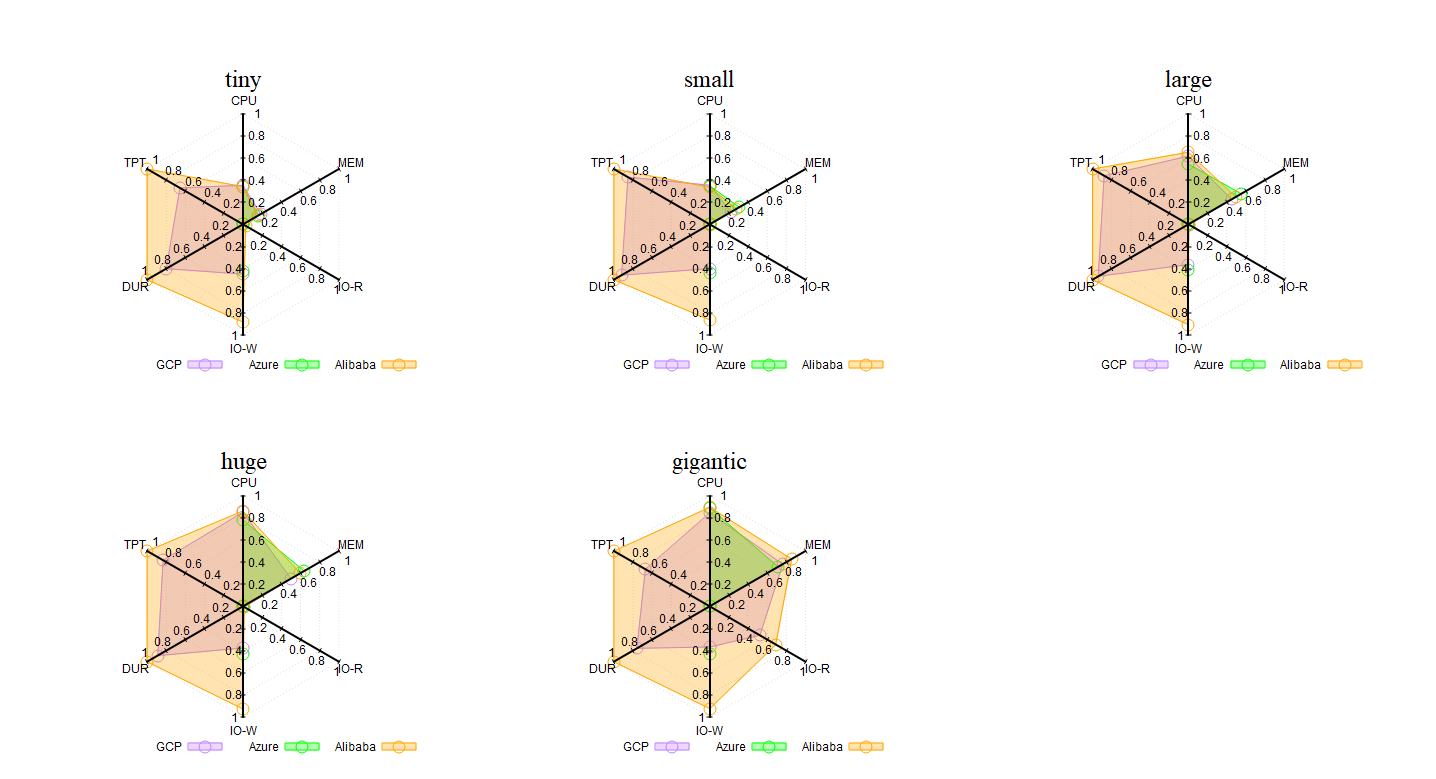}
 	\caption{Use Case 2 - Wordcount performances along with data scales}
 	\label{fig:uc2-wrdcnt-new}
 \end{figure}

The average cluster utilization plots for Sort benchmark in Figure \ref{fig:uc2-srt-new} indicate that HDInsight cluster is efficiently utilized when the data grows to Gigantic scale size. In the Wordcount benchmark (Figure \ref{fig:uc2-wrdcnt-new}), with high CPU utilization among the clusters, the I/O averages determined higher Throughput, hence, better Duration for e-MapReduce.

In order to develop a rationale for the resource utilization, Slowdown Estimates (SE), based on the raw benchmark results, are calculated for the second and third performances: \textit{SE = Duration / Shortest duration}. Table \ref{tab:uc1-comparative-results} groups the results within the data scales Huge and Gigantic. The column "Benchmark" refers to the workload; "First", contains the holder of the shortest duration; "Second" represents the service with the second shortest duration, "Third" lists the provider of the longest duration for the workload. Columns "SE" display the slowdown estimate of the provider listed to their left. In Table \ref{tab:uc2-comparative-results}, the first two columns represent the Workload and leveraged Data Scale, while the rest of the columns address the same as with Table \ref{tab:uc1-comparative-results}.

GCP Dataproc and Alibaba e-MapReduce services, in most cases, appeared to be showing similar utilization patterns in the linecharts. In parallel to the affinity, respective SE also indicate tiny differences as presented in Table \ref{tab:uc1-comparative-results}. With GCP's and Alibaba's similar architectures mentioned earlier, it is not surprising to observe likewise utilization behavior in both clusters.

On the other hand, Azure's HDInsight need to be evaluated from a different angle. There exist numerous samples, where HDInsight's SE values are notably higher. However, considering this as an inefficiency would lead us to a trap, because HDInsight's performance improvements over data scales would stay unnoticed. The architecture of HDInsight differs from the ones of Dataproc and e-MapReduce in that it is a cloud-optimized version of the Hortonworks Data Platform (HDP). In Table \ref{tab:uc1-comparative-results}, we observed HDInsight improving its SE significantly along the increasing data scale. For instance, in SQL category, its SE values improved from a range between 2.03 to 2.22 in Huge data scale, to a range between 1.14 to 1.28 in Gigantic data scale. For the Kmeans workload in the ML category, the SE value of HDInsight hits about 1.12 in huge data scale, and an insignificant SE in the Gigantic data scale. Also, for the Pagerank workload, its SE drops from 2.16 in Huge data scale to 1.41 in the Gigantic data scale.

The performance of Hadoop is affected by the mapper and reducer slot allocation capacity specified by the pre-configuration of managed system. Allocated mapper and reducer slots for various benchmarks are presented in Table \ref{tab:uc1-sql-mr-allocs} and Table \ref{tab:uc2-mr-allocs}. We gathered the values from the job counts within the MapReduce execution logs that were stored in HiBench reports. The allocation tables support our assumption that similar architectures (Dataproc and e-MapReduce) produce similar number of mapper slots during the map phases.

%%Use Case 1 benchmark results:
\begin{table}
	\centering
	\small
	\caption{Use Case 1 benchmark results }
	\label{tab:uc1-results}
	\begin{tabular}[!htb]{ l l r r r r r r r }
		\multicolumn{9}{ c }{Data Scale: Huge} \\
		{} & {} & {} & \multicolumn{2}{ c }{\textbf{Dataproc}} & \multicolumn{2}{ c }{\textbf{HDInsight}} & \multicolumn{2}{ c }{\textbf{e-MapReduce}} \\
		\hline
		{Category} & {Benchmark} & {IDS} & \begin{math}D_{(s)}\end{math} & \begin{math}T_{(MB/s)}\end{math} & \begin{math}D_{(s)}\end{math} & \begin{math}T_{(MB/s)}\end{math} & \begin{math}D_{(s)}\end{math} & \begin{math}{T_{(MB/s)}}\end{math} \\
		\hline
		\multirow{5}{ 1em }{Micro} & Sort & 3.28 & 70 & 47.11 & 131 & 25.08 & 111 & 29.42 \\
		{} & Terasort & 32.00 & 667 & 47.99 & 858 & 37.28 & 1054 & 30.37 \\
		{} & Wordcount & 32.85 & 978 & 33.60 & 1470 & 22.34 & 889 & 36.95 \\
		{} & Dfsioe-r & 26.99 & 294 & 91.77 & 662 & 40.79 & 245 & 110.21 \\
		{} & Dfsioe-w & 27.16 & 379 & 71.73 & 658 & 41.30 & 281 & 96.49 \\
		\hdashline
		\multirow{3}{ 1em }{SQL} & Scan & 2.01 & 73 & 27.63 & 157 & 12.83 & 74 (*) & 27.19 (*) \\
		{} & {Join} & 1.92 & 181 & 10.61 & 356 & 5.39 & 175 (*) & 10.95 (*) \\
		{} & Aggregation & 0.37 & 97 & 3.86 & 215 & 1.73 & 97 (*) & 3.85 (*) \\
		\hdashline
		\multirow{2}{ 1em }{ML} & Bayes & 1.88 & 2604 & 0.72 & 6120 & 0.31 & 3017 & 0.62 \\
		{} & Kmeans & 20.08 & 2321 & 8.65 & 2313 & 8.68 & 2070 & 9.70 \\
		\hdashline
		Websearch & Pagerank & 2.99 & 1544 & 1.94 & 3334 & 0.90 & 2458 & 1.22 \\
		\hline
		\multicolumn{9}{ c }{Data Scale: Gigantic} \\
		{} & {} & {} & \multicolumn{2}{ c }{\textbf{Dataproc}} & \multicolumn{2}{ c }{\textbf{HDInsight}} & \multicolumn{2}{ c }{\textbf{e-MapReduce}} \\
		\hline
		{Category} & {Benchmark} & {IDS} & \begin{math}D_{(s)}\end{math} & \begin{math}T_{(MB/s)}\end{math} & \begin{math}D_{(s)}\end{math} & \begin{math}T_{(MB/s)}\end{math} & \begin{math}D_{(s)}\end{math} & \begin{math}{T_{(MB/s)}}\end{math} \\
		\hline
		\multirow{5}{ 1em }{Micro} & Sort & 32.85 & 715 & 45.94 & 787 & 41.72 & 896 & 36.68 \\
		{} & Terasort & 320.00 & 9821 & 32.58 & ---(**) & ---(**) & 9660 & 33.13 \\
		{} & Wordcount & 328.49 & 10131 & 32.42 & 13596 & 24.16 & 8671 & 37.88 \\
		{} & Dfsioe-r & 216.03 & 915 & 236.11 & 1886 & 114.54 & 660 & 327.29 \\
		{} & Dfsioe-w & 217.33 & 1347 & 161.39 & 1914 & 113.57 & 1060 & 205.12 \\
		\hdashline
		\multirow{3}{ 1em }{SQL} & Scan & 20.10 & 457 & 43.96 & 514 & 39.09 & 407 (*) & 49.38 (*) \\
		{} & Join & 19.19 & 595 & 32.27 & 761 & 25.24 & 594 (*) & 32.32 (*) \\
		{} & Aggregation & 3.69 & 523 & 7.05 & 594 & 6.20 & 565 (*) & 6.52 (*) \\
		\hdashline
		\multirow{2}{ 1em }{ML} & Bayes & 3.77 & 5350 & 0.70 & 12589 & 0.30 & 6363 & 0.60 \\
		{} & Kmeans & 40.16 & 4541 & 8.84 & 4042 & 9.94 & 4034 & 9.96 \\
		\hdashline
		Websearch & Pagerank & 19.93 & 8371 & 2.38 & 11779 & 1.70 & 13893 & 1.43 \\
		\hline
		\multicolumn{9}{r}{IDS: Input Data Size (GB); \begin{math}D_{(s)}\end{math}: Duration (sec); \begin{math}T_{(MB/s)}\end{math}: Throughput (MB/sec)} \\
		\multicolumn{9}{ r }{(*) Benchmark succeeds after modifying preconfiguration, more on this in Discussion } \\
		\multicolumn{9}{ r }{(**) Workload failed to run within 3 attempts, discussed in Section \ref{Subsec:RESULTS_ISSUES_BOTTLENECKS} } \\
		\hline
	\end{tabular}
\end{table}

%% Use Case 2 benchmark Results:
\begin{table}
	\centering
	\small
	\caption{Use Case 2 benchmark results}
	\label{tab:uc2-results}
	\begin{tabular}[!htb]{ l l r r r r r r r }
		{} & {} & {} & \multicolumn{2}{ c }{\textbf{Dataproc}} & \multicolumn{2}{ c }{\textbf{HDInsight}} & \multicolumn{2}{ c }{\textbf{e-MapReduce}} \\
		\hline
		{Work} & {Scale} & {IDS} & \begin{math}D_{(s)}\end{math} & \begin{math}T_{(MB/s)}\end{math} & \begin{math}D_{(s)}\end{math} & \begin{math}T_{(MB/s)}\end{math} & \begin{math}D_{(s)}\end{math} & \begin{math}{T_{(MB/s)}}\end{math} \\
		\hline
		\multirow{5}{ 4em }{Sort} & Tiny & 39.30 KB & 36 & 0.0012 & 69 & 0.0006 & 32 & 0.0012 \\
		{} & Small & 3.28 MB & 36 & 0.09 & 70 & 0.0471 & 31 & 0.105 \\
		{} & Large & 328.50 MB & 42 & 7.86 & 81 & 4.07 & 42 & 7.74 \\
		{} & Huge & 3.28 GB & 70 & 47.08 & 141 & 23.36 & 107 & 30.69 \\
		{} & Gig. & 32.85 GB & 694 & 47.30 & 699 & 47.00 & 883 & 37.20 \\
		\hdashline
		\multirow{5}{ 4em }{Wrdcnt} & Tiny & 38.65 KB & 38 & 0.001 & 68 & 0.0006 & 31 & 0.0012 \\
		{} & Small & 348.29 MB & 50 & 6.51 & 98 & 3.34 & 47 & 7.06 \\
		{} & Large & 3.28 GB & 129 & 25.45 & 269 & 12.20 & 120 & 27.27 \\
		{} & Huge & 32.85 GB & 952 & 34.51 & 1487 & 22.10 & 888 & 37.00 \\
		{} & Gig. & 328.49 GB & 9749 & 33.70 & 13286 & 24.73 & 8622 & 38.10 \\
		\hline
		\multicolumn{9}{r}{IDS: Input Data Size; \begin{math}D_{(s)}\end{math}: Duration (sec); \begin{math}T_{(MB/s)}\end{math}: Throughput (MB/sec)} \\
		\hline
	\end{tabular}
\end{table}

%%Use Case 1 Slowdown Estimates:
\begin{table}
	\centering
	\small
	\caption{Use Case 1 Slowdown Estimates}
	\label{tab:uc1-comparative-results}
	\begin{tabular}[!htb]{ l l l r r r r }
		\multicolumn{7}{ c }{DATA SCALE: HUGE}  \\
		{Category} & {Benchmark} & First & Second & SE & Third & SE \\
		\hline
		\multirow{5}{ 1em }{Micro} & Sort & GCP & Alibaba & 1.59 & Azure & 1.87 \\
		{} & Terasort & GCP & Azure & 1.29 & Alibaba & 1.58 \\
		{} & Wordcount & Alibaba & GCP & 1.10 & Azure & 1.65 \\
		{} & Dfsioe-r & Alibaba & GCP & 1.20 & Azure & 2.70 \\
		{} & Dfsioe-w & Alibaba & GCP & 1.35 & Azure & 2.34 \\
		\hdashline
		\multirow{3}{ 1em }{SQL} & Scan & GCP & Alibaba & 1.01 & Azure & 2.15 \\
		{} & Join & Alibaba & GCP & 1.03 & Azure & 2.03 \\
		{} & Aggregation & GCP-Alibaba & Azure & 2.22 & --- & --- \\
		\hdashline
		\multirow{2}{ 1em }{ML} & Bayes & GCP & Alibaba & 1.16 & Azure & 2.35 \\
		{} & Kmeans & Alibaba & Azure-GCP & 1.12 & --- & --- \\
		\hdashline
		Websearch & Pagerank & GCP & Alibaba & 1.59 & Azure & 2.16 \\
		\hline
		\multicolumn{7}{ c }{DATA SCALE: GIGANTIC}  \\
		{Category} & {Benchmark} & First & Second & SE & Third & SE \\
		\hline
		\multirow{5}{ 1em }{Micro} & Sort & GCP & Azure & 1.10 & Alibaba & 1.25 \\
		{} & Terasort & Alibaba & GCP & 1.02 & Azure & (*) \\
		{} & Wordcount & Alibaba & GCP & 1.17 & Azure & 1.57 \\
		{} & Dfsioe-r & Alibaba & GCP & 1.39 & Azure & 2.86 \\
		{} & Dfsioe-w & Alibaba & GCP & 1.27 & Azure & 1.81 \\
		\hdashline
		\multirow{3}{ 1em }{SQL} & Scan & Alibaba & GCP & 1.12 & Azure & 1.26 \\
		{} & Join & Alibaba-GCP & Azure & 1.28 & --- & --- \\
		{} & Aggregation & GCP & Alibaba & 1.08 & Azure & 1.14 \\
		\hdashline
		\multirow{2}{ 1em }{ML} & Bayes & GCP & Alibaba & 1.19 & Azure & 2.35 \\
		{} & Kmeans & Alibaba-Azure & GCP & 1.13 & --- & --- \\
		\hdashline
		Websearch & Pagerank & GCP & Azure & 1.41 & Alibaba & 1.66 \\
		\hline
		\multicolumn{7}{ r }{(*) Workload failed to run within 3 attempts, discussed in Section \ref{Subsec:RESULTS_ISSUES_BOTTLENECKS}}  \\
		\hline
	\end{tabular}
\end{table}

%%Use Case 2 Slowdown Estimates:
\begin{table}
	\centering
	\small
	\caption{Use Case 2 Slowdown Estimates}
	\label{tab:uc2-comparative-results}
	\begin{tabular}[!htb]{ l l l r r r r }
		Work & {Scale} & First & Second & SE & Third & SE \\
		\hline
		\multirow{5}{ 4em }{Sort} & Tiny & Alibaba & GCP & 1.13 & Azure & 2.16 \\
		{} & Small & Alibaba & GCP & 1.16 & Azure & 2.26 \\
		{} & Large & GCP-Alibaba & Azure & 1.93 & --- & --- \\
		{} & Huge & GCP & Alibaba & 1.53 & Azure & 2.01 \\
		{} & Gig. & GCP & Azure & 1.01 & Alibaba & 1.27 \\
		\hdashline
		\multirow{5}{ 4em }{Wrdcnt} & Tiny & Alibaba & GCP & 1.23 & Azure & 2.19 \\
		{} & Small & Alibaba & GCP & 1.06 & Azure & 2.09 \\
		{} & Large & Alibaba & GCP & 1.08 & Azure & 2.24 \\
		{} & Huge & Alibaba & GCP & 1.07 & Azure & 1.67 \\
		{} & Gig. & Alibaba & GCP & 1.13 & Azure & 1.54 \\
		\hline
	\end{tabular}
\end{table}

%% Allocated maps and reduces in SQL category
\begin{table}
	\centering
	\small
	\caption{Allocated map and reduce slots from Use Case 1 SQL category}
	\label{tab:uc1-sql-mr-allocs}
	\begin{tabular}[!htb]{ l l r r r r r r  }
		{} & {} & \multicolumn{2}{c}{GCP} & \multicolumn{2}{c}{Azure} & \multicolumn{2}{c}{Alibaba} \\
		{Work} & Scale & {Maps} & Reduces & {Maps} & Reduces & {Maps} & Reduces \\
		\hline
        \multirow{2}{ 4em }{Scan*} & Huge & 24 & -- & 12 & -- & 24 & --  \\
		{} & Gigantic & 144 & -- & 36 & -- & 144 & -- \\
		\hdashline
		\multirow{2}{ 4em }{Join} & Huge & 60 & 25 & 48 & 25 & 60 & 25 \\
		{} & Gigantic & 180 & 25 & 72 & 25 & 180 & 25 \\
		\hdashline
		\multirow{2}{ 1em }{Aggreg.} & Huge & 24 & 12 & 12 & 12 & 24 & 12 \\
		{} & Gigantic & 144 & 12 & 36 & 12 & 144 & 12 \\
		\hline
		\multicolumn{8}{ r }{* No reduce operation} \\
		\hline
	\end{tabular}
\end{table}

%% Allocated maps and reduces in Use Case 2
\begin{table}
	\centering
	\small
	\caption{Allocated map and reduce slots in Use Case 2}
	\label{tab:uc2-mr-allocs}
	\begin{tabular}[!htb]{ l l r r r r r r }
		{} & {} & \multicolumn{2}{c}{GCP} & \multicolumn{2}{c}{Azure} & \multicolumn{2}{c}{Alibaba} \\
		{Work} & {Scale} & {Maps} & Reduces & {Maps} & Reduces & {Maps} & Reduces \\
		\hline
		\multirow{5}{ 4em }{Sort} & Tiny & 11 & 12 & 12 & 12 & 11 & 12 \\
		{} & Small & 11 & 12 & 12 & 12 & 11 & 12 \\
		{} & Large & 12 & 11 & 12 & 12 & 11 & 11 \\
		{} & Huge & 24 & 12 & 12 & 12 & 23 & 12 \\
		{} & Gigantic & 251 & 12 & 60 & 12 & 252 & 12 \\
		\hdashline
		\multirow{5}{ 4em }{Wrdcnt} & Tiny & 11 & 12 & 12 & 12 & 11 & 12 \\
		{} & Small & 12 & 11 & 12 & 12 & 12 & 11 \\
		{} & Large & 24 & 11 & 12 & 12 & 23 & 12 \\
		{} & Huge & 250 & 11 & 60 & 12 & 252 & 12 \\
		{} & Gigantic & 2447 & 12 & 612 & 12 & 2447 & 11 \\
		\hline
	\end{tabular}
\end{table}

%[REVİZE: Başlıktaki "Performance Issues...", "Cluster Utilization Issues...." şeklinde değiştirildi - emre]

\subsection{Cluster Utilization Issues and Bottlenecks}\label{Subsec:RESULTS_ISSUES_BOTTLENECKS}

This section deals with issues that occurred in two workloads; Bayes and Kmeans in Gigantic data scales in Use Case 1.

\begin{figure}[!htb]
	\includegraphics[width=\textwidth]{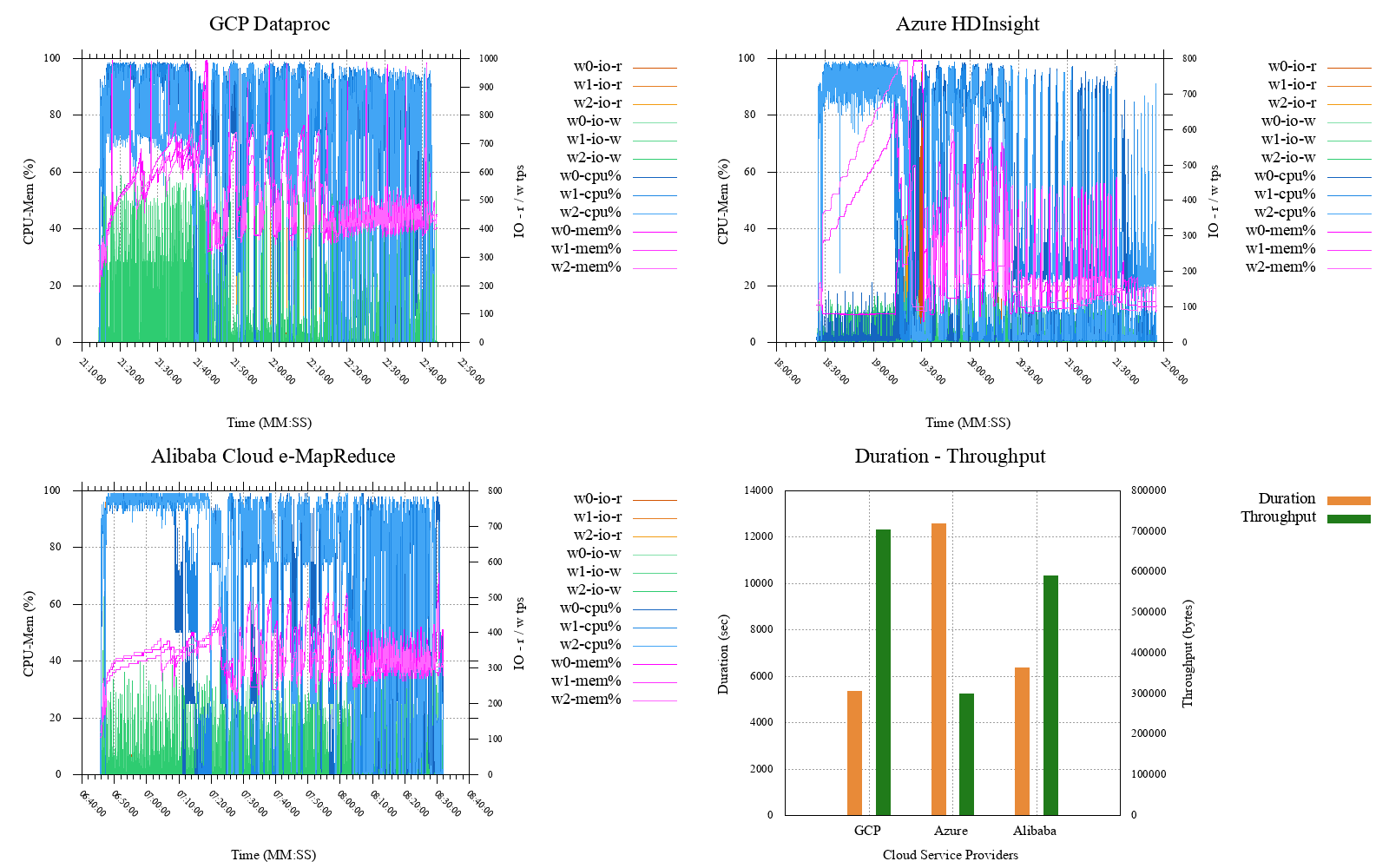}
	\caption{Use Case 1 - Bayes (Gigantic; PAGES: 1,000,000 CLASSES: 100 NGRAMS: 2)}
	\label{fig:uc1-bayes-g-cmidt}
\end{figure}

In Table \ref{tab:uc1-comparative-results}, even though HDInsight's cluster utilization for the Kmeans workload within the data scales Huge and Gigantic were resulting in promising SE, this was not the case for the Bayes workload. This is contradictory since both workloads leverage the Mahout engine. Figure \ref{fig:uc1-bayes-g-cmidt} plots the difference of HDInsight's Throughput and Duration compared to other services for the Bayes workload. The linechart for HDInsight displays low IO transfers and fluctuations in CPU and Memory. While we were investigating the MapReduce execution logs, it turned out that a Java Heap space error caused the process to restart nine times, occuring at about 83 percent completion rate within each Map process, thus prolonging the overall duration of the benchmark. Considering that the map JVM heap size is a property (\textit{mapreduce.map.java.opts}) that could be set within the configurations, we see this occurence as an insufficiency in managed Hadoop context's default pre-configuration that needs to be further improved. The MapReduce execution logs for Bayes workload are available at the referenced URL \cite{docu_bayes_logs}. 

\begin{figure}[!htb]
 	\includegraphics[width=\textwidth]{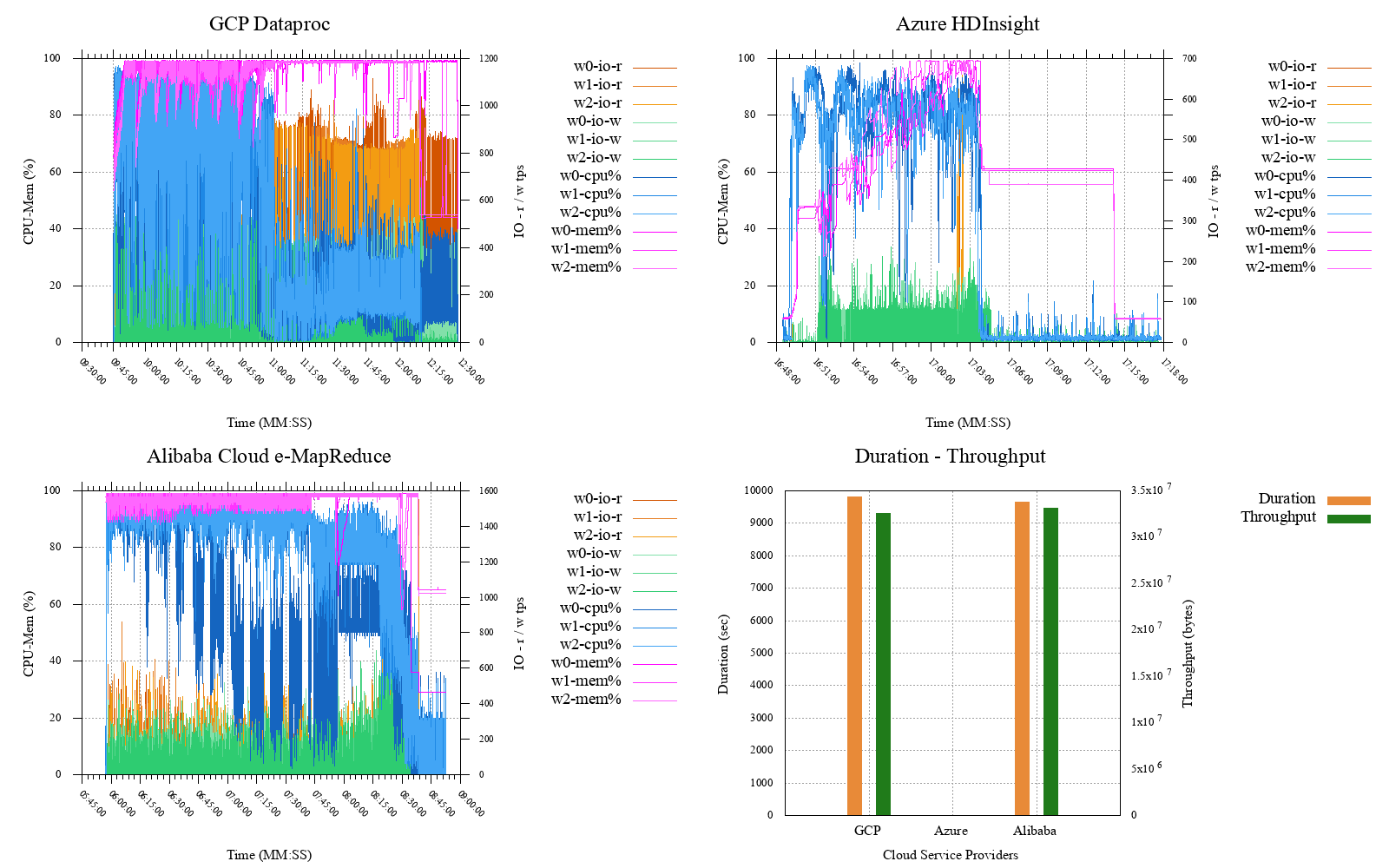}
 	\caption{Use Case 1 - Terasort (Gigantic; 3.2 GB)}
 	\label{fig:uc1-tera-g-cmidt}
 \end{figure}

In the second case, we inspect the Terasort benchmark running in data scale Gigantic for HDInsight. Throughout the three attempts we conducted, the execution was terminated when around 20 percent of the map process was reached. On Dataproc and e-MapReduce, the end-user was allowed to choose HDFS or the respective provider's cloud storage service as storage option, and also possessed the option to specify the size of the local disk for the worker nodes. In HDInsight, the end-user was allowed to leverage WASB blob storage. However, as there was no option to specify a local HDFS disk size for the worker nodes, a DiskErrorException occurred. Meaning, there was no sufficient place on the local disk of the worker node, where intermediate map results could be written. We marked this shortcoming as a structural bottleneck since all end users who run Terasort operation at this scale would face the same error. Figure \ref{fig:uc1-tera-g-cmidt} displays system resource utilization on HDInsight, including the point where the failure occurs and becomes a flatline. Due to the incomplete execution, Duration and Throughput results for HDInsight could not be provided in the graphicst. The error logs are available at the referenced URL for the interested \cite{docu_terasort_logs}.

\section{Discussion}\label{discuss}
In the study, we investigated the utilization behavior of the managed Hadoop context by benchmarking experimental environments from different providers. For a fair assessment, we activated the services within same geographical data center locations, same computation specifications. Minor differences in Hadoop version did occur, and no configuration tweak is applied to the clusters. 

%[REVİZE: benchmark'ların 3'er kere çalıştırılmaması bir kısıtlamanın sonucu olarak değil, ama bir tercih olarak ifadelendirildi. Amacı en hızlı provider'ı tespit etmek olmayan, ama kaynak kullanım davranışlarına odaklanmak isteyen bu çalışmada 3 benchmark'ın ortalamasını almanın marjinal fayda olarak katkısı çok düşük - emre] 

%Within the constraints of an experimental study, we had to limit benchmark executions at only one successful execution for each workload. In future work, we plan on executing each benchmark three times and taking the average, which would provide more stable outcomes, especially in cases, where there were small differences between subsequent performances of respective providers. 

A general approach in benchmark is to run a specific workload multiple (mostly three) times and take the average result as its final value. This is crucial for performance-focused or commercially driven benchmarks. We deviated from this approach by one execution per benchmark; the rationale is drawn by our interest to stand observative on cluster resource utilization in a non-performative boundary. Rather than taking average of multiple runs, we collected utilization data on each of the worker nodes during the runtime, which enabled us to plot cluster utilization over a time span. Performance results or Slowdown Estimates are visited as secondary indicators, mostly to observe the impact of the respective utilization behavior. The marginal benefit of timely-based average utilization data to the study's objective would be less significant. Our main concern was not to compare performances, but to understand inner dynamics within managed Hadoop context in experimental environments, hence, the decision.

%[REVİZE: benzer şekilde, HiBench'in predefined data scale'leri içinde en büyüğü olan Bigdata ölçeğini devre dışında tutmuş olsak bile 320GB'lık Wordcount örneği production için kabul edilebilir bir değer. Alttaki cümle metinden çıkarıldı - emre]

%Another limitation of the study was that we had to opt-out the largest predefined data scale of HiBench, namely Bigdata. The dataset sizes within this scale would go beyond the specified local storage capacity.

HiBench comes with dependencies downloaded during its compilation process by Apache Maven. The Hive engine is one of those dependencies leveraged by HiBench for running SQL workloads Scan, Join, and Aggregation. Alibaba e-Mapreduce comprised a ready-made Hive hook triggering a Java file to run post-executional transactions for other services within the package. This configuration prevented HiBench from starting SQL benchmarks since the HiBench based Hive engine did not include the jar file mentioned above, defined for e-MapReduce's specific environment. A workaround attempt, copying the related jar file to an appropriate directory within HiBench, made the jar file available. However, this time, HiBench's Hive engine of an older version did not support the hook "hive.exec.post.hooks" defined in Alibaba's Hive configuration. At this point, disabling the respective Hive hook from Alibaba e-MapReduce's UI management console apparently solved this issue and enabled HiBench's SQL workloads to run. Its impact on the respective system utilization remains unknown, hence the need to annotate it here. With GCP and Azure, issues of this kind did not occur.

\section{Conclusion and Future Work}
In the study, we conducted HiBench benchmarks against Hadoop PaaS clusters in their proposed, out-of-the-box form, provided by respective CSPs. For each service we specified the same geographical location, and apparently the same CPU and Memory properties as promised by the providers. 

Our aim was to inspect resource utilization within managed Hadoop context by leveraging HiBench Benchmark Suite. We executed Hadoop related benchmarks on three managed services and captured system resource utilization data on the worker nodes. 

Based on the benchmark results, we presented slowdown estimates to observe how the performances differ from each other. For selected benchmarks, we gathered the numbers of the allocated map and reduce slots by vendors' managed Hadoop systems. We found that there is a strong relation between sufficiently allocated map slots and cluster utilization, thus, their performances. It appears that default pre-configuration settings do not let sufficient resources to be utilized, and affecting their performance.

The results yielded that managed Hadoop services, even when defined with same properties, do not necessarily utilize system resources in similar ways. It is rather dependent to the managed Hadoop architecture, as we have observed how similar architectures followed a resembling pattern in utilization over time whereas a different architectural approach differs in its performance and utilization behavior. 

Managed Hadoop context do provide a comfort zone by means of automated implementation processes; however, as discussed in the study, this can manifest itself as a shortcoming on the respective system's resource utilization dynamics. By reading our study, the end-user would know managed systems are not magical instruments, and still require to investigate the respective managed system's architecture and implement performance increasing operations on them.

This study was intended as the first part of a series examining managed service context. Our future work will inspect further offerings of the managed context, such as Apache Spark, and enrich the scope with additional cloud vendors.

\end{document}